# Modeling Self-Similar Traffic for Network Simulation

Xiaofeng Bai, Abdallah Shami
Electrical and Computer Engineering Department
The University of Western Ontario

In order to closely simulate the real network scenario thereby verify the effectiveness of protocol designs, it is necessary to model the traffic flows carried over realistic networks. Extensive studies [1] showed that the actual traffic in access and local area networks (e.g., those generated by ftp and video streams) exhibits the property of *self-similarity* and *long-range dependency* (LRD) [2]. In this appendix we briefly introduce the property of self-similarity and suggest a practical approach for modeling self-similar traces with specified traffic intensity.

## I. What Is Self-Similarity ?

Self-similarity is also called *infinite variance syndrome* [3]. Simply speaking, a process shows self-similarity implies the process is indistinguishable from its scaled versions obtained by averaging the original process within different observation time scales. Missing this property makes some other popular traffic model such as *Poisson* trace give over-optimized evaluation of network performances. Figures 1a and 1b illustrate the different scaling behavior of self-similar and non-self-similar traffics.

A mathematical description of self-similarity can be concluded as follows [4]. Assume an increment process $X_i$ $(i = 1, 2, \cdots)$ and another process $X_j^{(m)}$ $(j = 1, 2, \cdots)$ which is obtained by averaging the values in non-overlapped blocks of size *m* in $X_i$, i.e.,

$$X_j^{(m)} = \frac{1}{m}\left(X_{jm-m+1} + X_{jm-m+2} + \cdots + X_{jm}\right) \qquad (1)$$

The process $X_i$ is said self-similar if

$$X_j^{(m)} \overset{dis}{\sim} m^{H-1} X_i \tag{2}$$

The symbol $\overset{dis}{\sim}$ denotes equality in distribution (note this does not mean the exactly same picture repeats) and the similarity between the distribution of $X_j^{(m)}$ and the distribution of $X_i$ decays by a power law, i.e., factor $m^{H-1}$. In a more understandable way, this implies

$$Var(X_j^{(m)}) = m^{2H-2} Var(X_i) \tag{3}$$

where $Var(\cdot)$ denotes the variance of a process. Here $m$ $(m \geq 1)$ is the *scale parameter*, whereas $H$ $(0.5 < H \leq 1)$ is the *Hurst parameter*. For an instance, when $H = 1$ process $X_j^{(m)}$ and process $X_i$ have the same distribution without any decay. In this case $X_i$ is also called second-order self-similar [3]. The Hurst parameter is used to measure the burstiness of a process.

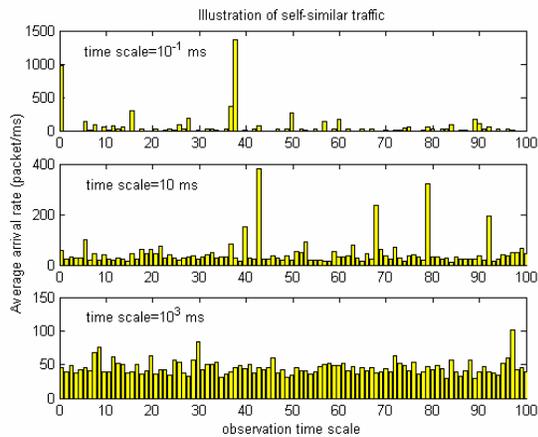 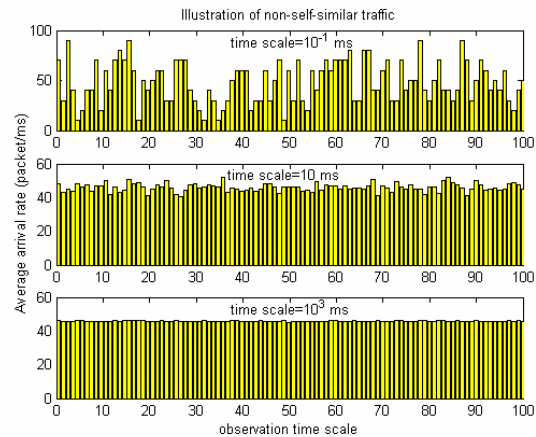

Fig. 1a: Illustration of self-similarity    Fig. 1b: Illustration of non-self-similarity

The LRD (long rang dependency) implies a non-summable *autocorrelation function* (ACF) of the process. It is proved in [2] when $(0.5 < H < 1)$ the ACF of $X_i$ is not summable. Accordingly, when $(0 < H < 0.5)$ $X_i$ is called *short range dependency* (SRD). Figures 2a and 2b show the ACFs of LRD and SRD traffics generated in our simulations. We can see that when the sequence is shifted, the ACF of SRD traffic converges to zero rather smoothly, while in the ACF of LRD traffic, very strong spikes appear in a wide area, i.e., long rang dependency.

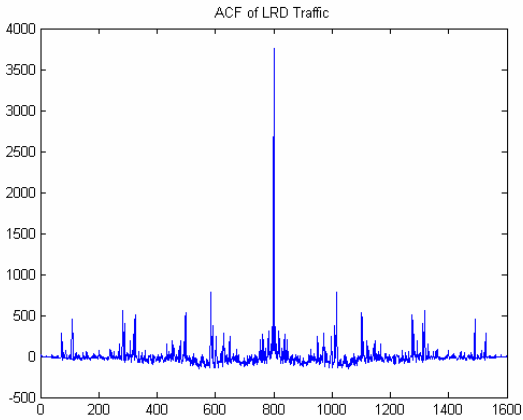 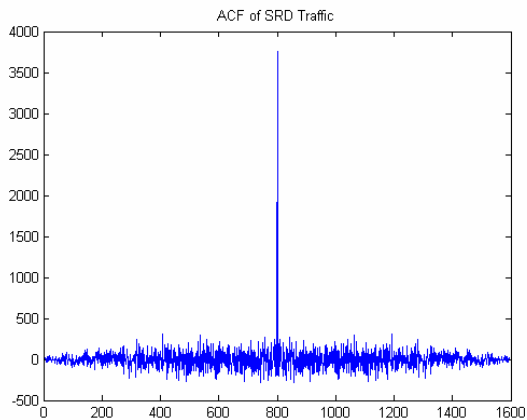

Fig. 2a: Illustration of LRD ACF          Fig. 2b: Illustration of SRD ACF

If we apply logarithm to both sides of eq. (3) we will have

$$\log\left(\frac{Var(X_j^{(m)})}{Var(X_i)}\right) = (2H - 2)\log(m) \qquad (4)$$

This implies that the log-log plot of $Var(X_j^{(m)})$ normalized by $Var(X_i)$, vs. the scale parameter *m*, will give a linear figure with slope of $2H - 2$. In other words, the Hurst parameter can be evaluated as $H = 1 + k/2$, where *k* is the linear slope in the log-log plot [2, 4]. Therefore, for LRD there is $-1 < k < 0$ (as $0.5 < H < 1$). Figure 3a shows the log-log plot of LRD and SRD traffics. In the figure we can see the log-log plot for LRD traffic does not show a strict linear slope. This is because of the "tail-truncation" effect; this effect will be discussed later. In Figure 3b, we approximate the slope as -0.6 using *least square* approach. Hence, the Hurst parameter is $H = 0.7$.

## II. Generating Self-Similar Traffic with Specified Intensity

The self-similarity is resulted from high probability for very large values that cannot be neglected. This can be understood as that a large packet burst makes the averaged value in a large observation time scale differentiated from its contemporaries, as shown in Figure 4. Therefore, when the large values occur so often that their effect cannot be neglected, the variance of the trace decays very slow while the observation scale increases, i.e., exhibiting self-similarity. Therefore, to model self-similar trace we need some distribution with "heavy-tailed" *probability density function* (PDF).

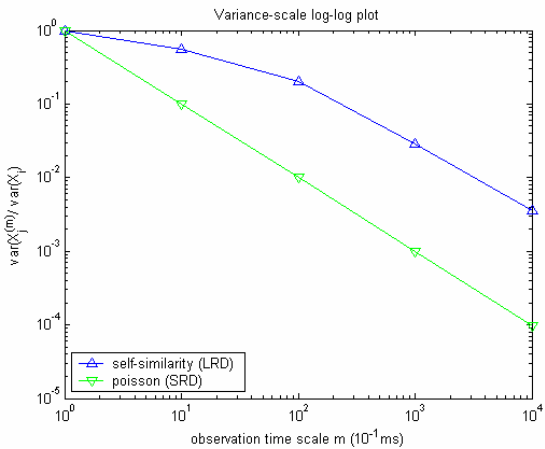
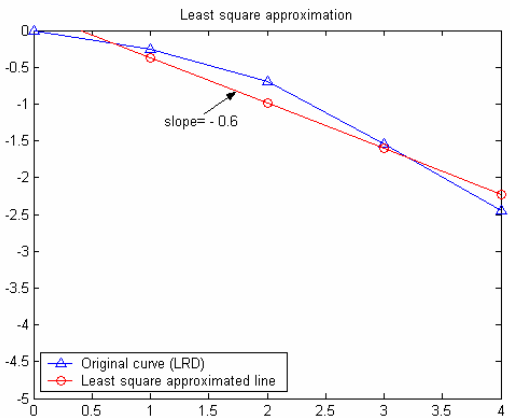

Fig. 3a: Log-log plot of LRD and SRD traffics    Fig. 3b: Least square approximation of the slope

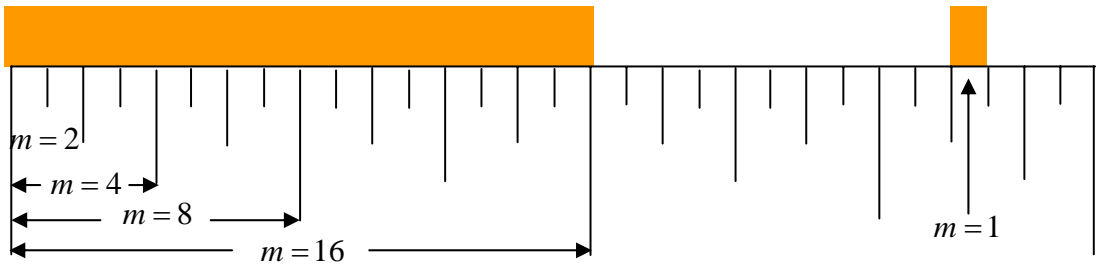

Fig. 4: Illustration of the large value effect

In practice we usually choose *Pareto distribution* to model "heavy-tailed" traffic trains. The PDF of Pareto distribution shown in Figure 5 is presented as:

$$P(x) = \frac{\alpha \beta^\alpha}{x^{\alpha+1}} \quad x \geq \beta \tag{5}$$

where $\alpha$ is the *shape parameter* and $\beta$ is the minimum value of $x$. The mean and variance of this distribution are:

$$\mu = \frac{\alpha \beta}{\alpha - 1} \tag{6}$$

and

$$\sigma^2 = \frac{\alpha \beta^2}{(\alpha - 1)^2 (\alpha - 2)} \tag{7}$$

We can see that when $\alpha > 1$ this distribution has finite mean, whereas when $\alpha < 2$ it has infinite variance. Therefore, to model self-similar traffic we need $1 < \alpha < 2$. To generate a sequence of values following Pareto distribution we can apply the *inverse CDF* (ICDF) method as [5]:

$$x = \frac{\beta}{s^{1/\alpha}} \tag{8}$$

where $s$ is a uniformly distributed value with $s \subset (0, 1]$.

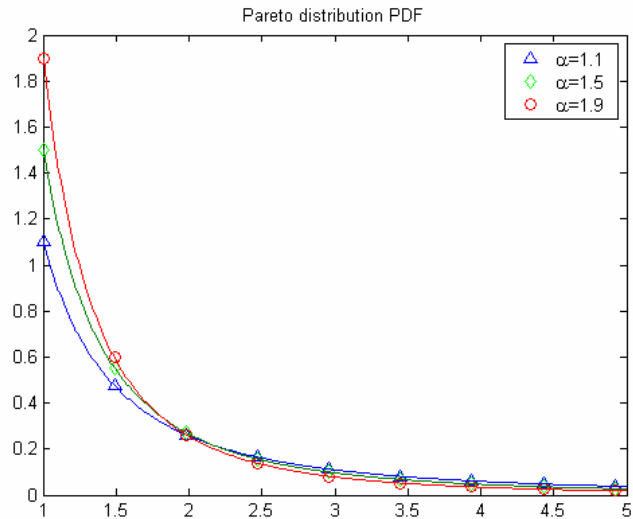

Fig. 5: PDF of Pareto distribution

Though Pareto distribution exhibits heavy-tail property, the values that can be practically generated in simulations only follow a truncated version of Pareto distribution, where the largest value is a finite number. This is raised by two reasons. First, from eq. (8) we can see the largest value that can be generated by computer is decided by the smallest $s$, which is a fixed value. This determines that the generated trace is always tail-truncated. Second, the number of values generated in simulation is finite. This indicates the tail-truncation may be even worse. If the smallest $s$ is $2^{-32}$, to obtain the least tail-truncation, $2^{32}$ values should be generated. In general, the tail-truncation effect raised by limited number of data used in simulation is more severe than the least tail-truncation the computer can offer. Consider Figure 4, this severe tail-truncation makes the decay of variance for large observation scale become faster than expected. This explains the non-linear slope in Figure 3a.

In our simulations, we applied the approach introduced in [3] to model self-similar traffics. This approach stated that the aggregation of multiple streams with strict alternating ON and OFF periods can result in a self-similar trace. In each of these streams, the ON period denotes a packet burst with the packet number following Pareto distribution, whereas the OFF period indicates a "silent" time interval with the duration following another separate Pareto distribution. The packet size can be the same within one stream but different from other streams. In our simulation we apply packet size uniformly distributed between 64 and 1518 bytes, which are the minimum and maximum sizes of Ethernet packets, for each stream.

Considering the tail-truncation effect occurred in practical simulation, the mean value of the obtained Pareto distribution is then computed as:

$$E(x) = \int_{\beta}^{\omega} xP(x)dx = \int_{\beta}^{\omega} x\frac{\alpha\beta^{\alpha}}{x^{\alpha+1}}dx = \frac{\alpha\beta}{\alpha-1}\left[1-\left(\frac{\beta}{\omega}\right)^{\alpha-1}\right] \qquad (9)$$

where $\omega$ denotes the largest value generated in a specific simulation run. This value is contingent on the number of ON and OFF periods occurred in the simulation. Namely, the more packets (more bursts) generated, the closer of the value computed by eq. (6) to the value computed by eq. (9). However, in simulation we only generate limited number of packets. Therefore, if the parameters are set based on eq. (6), the average bit rate offered by the generated packets is always different from the specified load. As shown in Figure 6, there is an error that converges to zero while the generated number of packets increases. In the following discussion we develop an approach to control this error within acceptable level for the limited number of packets generated, by choosing appropriate parameters.

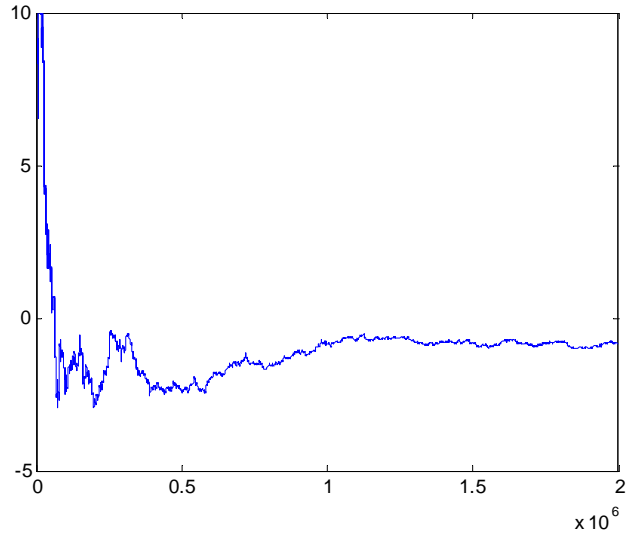

Fig. 6: Illustration of self-similar traffic generation error

Based on the approach for modeling self-similar traffic introduced before, we have:

$$r = \frac{E(b_{on})}{E(t_{on}) + E(t_{off})} \times N \qquad (10)$$

where $r$ is the specified bit rate (load), $b_{on}$ denotes the number of offered bits in an ON-period, $t_{on}$ and $t_{off}$ denote the time interval of an ON and an OFF period, respectively, and $N$ is the number of ON-OFF sources that aggregate the self-similar traffic. Applying eq. (6) to eq. (10), we have:

$$r = \frac{\left[\alpha_{on}\beta_{on}/(\alpha_{on}-1)\right]\left[(S_p + S_o)\times 8\right]}{\left[\alpha_{on}\beta_{on}/(\alpha_{on}-1)\right]\left[(S_p + S_o)\times 8\right]/R + \alpha_{off}\beta_{off}/(\alpha_{off}-1)} \times N \qquad (11)$$

where $\alpha_{on}$, $\beta_{on}$, $\alpha_{off}$, and $\beta_{off}$ denote the corresponding parameters in Pareto distribution for the ON-period and OFF-period of an ON-OFF source, $R$ is the bit rate of the link to which the self-similar trace is offered, and $S_p$ and $S_o$ denote the size in bytes of a data packet and the size in bytes of the packet tax (i.e., 96-bit IPG and 64-bit preamble in front of each packet). $S_p$ is 791 since the size of packet at each ON-OFF source is uniformly distributed between 64 and 1518 bytes. Eq. (11) can be rewritten as:

$$\frac{\beta_{on}}{\beta_{off}}\left(\frac{N(S_p+S_o)\times 8}{r}-\frac{(S_p+S_o)\times 8}{R}\right)=\frac{\alpha_{off}(\alpha_{on}-1)}{\alpha_{on}(\alpha_{off}-1)} \qquad (12)$$

Since there is tail-truncation effect, if the parameters are chosen according to eq. (12) the offered bit rate by limited number of packets generated in simulation does not equal to $r$. In eq. (12) only $\alpha_{on}, \alpha_{off}, \beta_{off}$ and $N$ is adjustable, as $\beta_{on}=1$ means the minimum burst size is one packet, which is always the truth. From Figure 5 we can see the tail-truncation effect can be partially adjusted by changing the value of $\alpha$ (i.e., the extent of heavy-tailness). Though changing $\alpha$ (but keep $1<\alpha<2$) affects the burstiness of the generated trace, by this we can enforce the error between generated bit rate (by limited number of packets) and specified load into an acceptable (specified) percentage. In other words, changing the value of $\alpha$ indirectly adjusts the reliability of eq. (6) and eq. (12) for evaluating the generated bit rate. As we also know increasing $N$ or decreasing $\beta_{off}$ will augment the generated bit rate, and vice versa. Our principle is to use these two variables to adjust other parameters (i.e., $\alpha_{on}$ and $\alpha_{off}$) such that the generated bit rate by limited number of packets meets the specified load with acceptable error.

Denoting the left part of eq. (12) by $\phi$, to decide the value of $\alpha_{on}$ and $\alpha_{off}$, from eq. (12) we have:

$$\alpha_{off} = \frac{\phi}{1/\alpha_{on} + \phi - 1} \tag{13}$$

Considering $1 < \alpha_{on} < 2$ and $1 < \alpha_{off} < 2$, the value of $\alpha_{on}$ should be:

$$\begin{cases} 1 < \alpha_{on} < 2 & \phi \geq 2 \\ 1 < \alpha_{on} < Min\left(2, \frac{1}{1-\phi/2}\right) & \phi < 2 \end{cases} \tag{14}$$

From eq. (14) we noticed that the value of $\alpha_{on}$, which mainly decides the burstiness, is not necessarily changed for every adjustment. Now the error control between generated bit rates and specified loads can be done by appropriately choosing $N$ and $\beta_{off}$. We propose an asymptotical adjustment approach to enforce the above error into acceptable range by repeating the self-similar trace generation with different values of $N$ and $\beta_{off}$. The background basis of this approach is to use practically resultant error to adjust the source number $N$ and minimum gap $\beta_{off}$. After appropriate parameters are set, the data is generated and output into file. This approach is explained with more details in the flow chart of Figure 7.

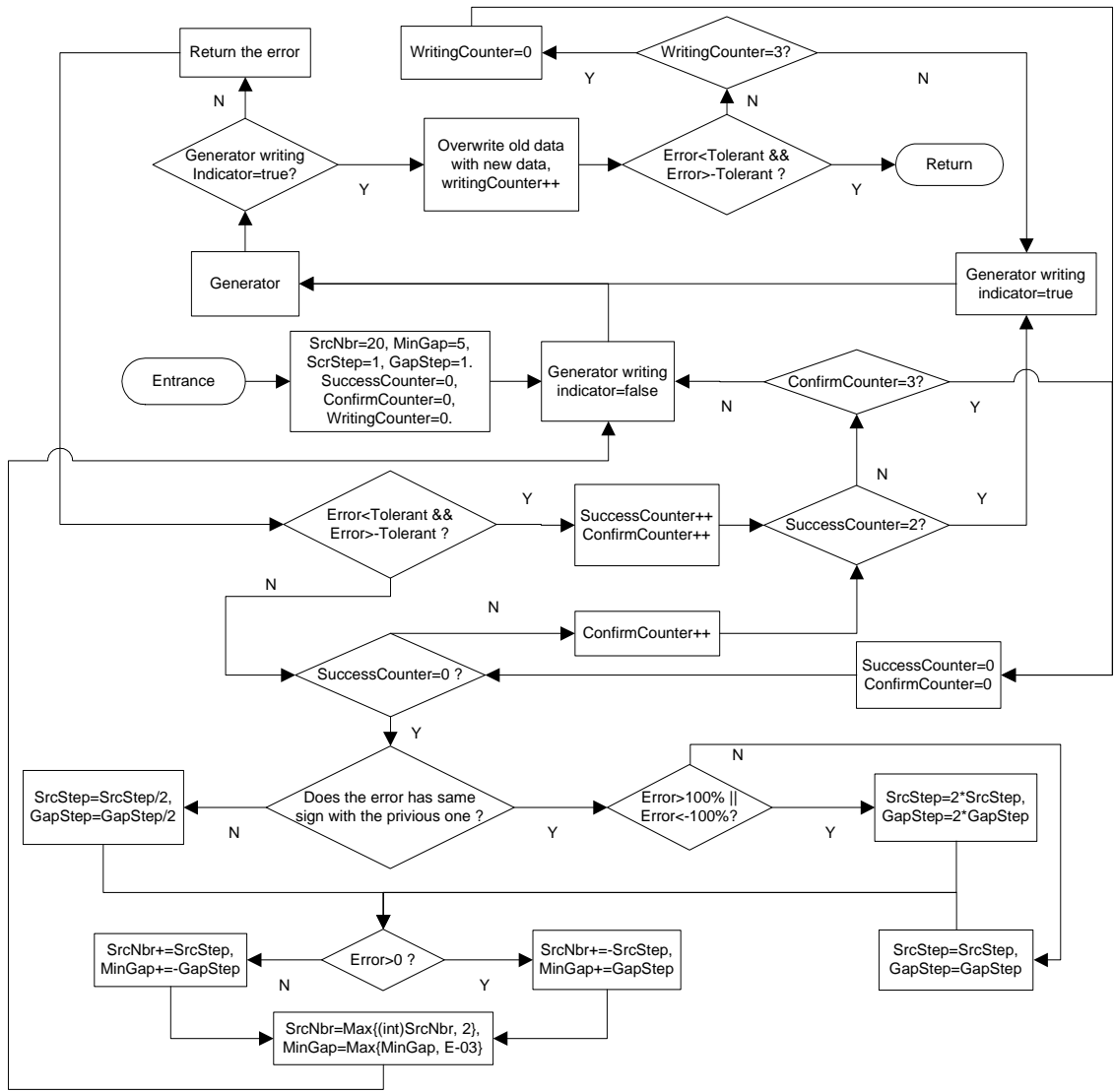

Fig. 7: Self-similar traffic generator error control flow chart